\documentclass[pra,onecolumn,epsfig,superscriptaddress,showpacs]{revtex4-2}
\usepackage{graphicx}
\usepackage{subcaption}
\makeatletter
\long\def\@makecaption#1#2{%
  \vskip\abovecaptionskip
  \sbox\@tempboxa{#1: #2}%
  \ifdim \wd\@tempboxa >\hsize
    #1: #2\par
  \else
    \global \@minipagefalse
    \hb@xt@\hsize{\hfil\box\@tempboxa\hfil}%
  \fi
  \vskip\belowcaptionskip}
\makeatother
\usepackage{epstopdf}
\usepackage{amsfonts}
\usepackage{amssymb}
\usepackage{amsmath}

\usepackage{prettyref}
\usepackage{float}
\usepackage{amsmath}
\usepackage{amssymb}
\usepackage{esint}
\usepackage{qcircuit}
\usepackage{color}
\usepackage{CJK}

\usepackage{braket}
\usepackage[left=1in, right=1in, top=1in, bottom=1in]{geometry}
\usepackage{tikz}
\usetikzlibrary{shapes.geometric, arrows}
\def\be{\begin{equation}}
\def\ee{\end{equation}}
\def\ba{\begin{eqnarray}}
\def\ea{\end{eqnarray}}

\def\pnp{${\text P}^{\sharp \text{P}}$ }

\begin{document}
\begin{CJK*}{UTF8}{gbsn}
\title
{Physics and computation: An insight from non-Hermitian quantum computing}

\author{Qi Zhang(张起)}
\affiliation{College of Science, Liaoning Petrochemical University,
Fushun 113001, China}

\affiliation{Liaoning Provincial Key Laboratory of Novel Micro-Nano Functional Materials,
Fushun 113001, China}

\author{Biao Wu(吴飙)}
\affiliation{International Center for Quantum Materials, Peking University, 100871, Beijing, China}
\affiliation{Wilczek Quantum Center, Shanghai Institute for Advanced Studies, Shanghai 201315, China}
\affiliation{Hefei National Laboratory, Hefei 230088, China}

\begin{abstract}
We elucidate the profound connection between physics and computation by proposing and examining the model of the non-Hermitian quantum computer (NQC). In addition to conventional quantum gates such as the Hadamard, phase, and CNOT gates, this model incorporates a non-unitary quantum gate $G$. We show that NQC is extraordinarily powerful, capable of solving not only all NP problems but also all problems within the complexity class $\text{P}^{\sharp\text{P}}$ in polynomial time. We investigate two physical schemes for implementing the non-unitary gate $G$ and find that the remarkable computational power of NQC originates from the exponentially large physical resources required.
\end{abstract}

\maketitle

\section{Introduction}
The interplay between physics and computation is evident in two key aspects: (1) the choice of computing medium
and (2) the physical principles underlying a computing model. A prime example of the former
is the transition from vacuum-tube-based computers to transistor-based ones.
The latter is exemplified by quantum computers~\cite{Nielson}, where one can execute more
powerful algorithms which are  infeasible in principle for classical computers,
such as Shor's algorithm~\cite{Shor} and Grover's algorithm~\cite{Grover}.

However, to date, the known computational power of quantum computers remains relatively limited,
and they are not yet able to efficiently solve NP-complete problems.
To achieve stronger computational power, various theoretical computing models have been proposed
by extending physical principles beyond standard quantum mechanics,
such as utilizing closed timelike curves~\cite{Deutsch,Bacon}, incorporating nonlinear gates~\cite{Abrams},
exploiting postselection~\cite{aaronson,knill}, and using Lorentzian gates~\cite{He,ZhangWu}.
All of these models can solve NP-complete problems in polynomial time; some of them
can even solve problems in the time complexity classes of PSPACE~\cite{aaronson1,aaronson2} and  \pnp~\cite{ZhangWu}.

In this work, we explore the profound relation between computation and physics by
proposing another theoretical computing model that is capable of
solving NP-complete problems in polynomial time. In this model, we relax the unitarity constraint
by introducing non-unitary gates. We demonstrate that this computing model consists of four universal gates:
Hadamard gate ($H$), phase gate ($T$), CNOT gate and non-unitary $G$ gate.
We define the class of problems solvable by non-Hermitian quantum
computers (NQC) in polynomial time as bounded-error non-Hermitian quantum polynomial time (BNQP)
and show that BNQP is equivalent to the complexity class ${\text P}^{\sharp \text{P}}$,
similar to the Lorentz quantum computer~\cite{ZhangWu}.
The power of this non-Hermitian computing model is then illustrated with an algorithm
that solves NP-hard problems in polynomial time. Through this concrete example,
it is evident that the power of this computing model comes from the non-unitary $G$ gate.

However, there is a crucial distinction between the current model and all of the previously
discussed models that are more powerful than conventional quantum computers:
the non-unitary gate $G$ can be implemented using real physical systems, specifically non-Hermitian systems.
These systems, which were theoretically proposed in the study of PT-symmetric systems and exceptional topology~\cite{Bender,BergholtzRMP,Alvarez,Meden,Gao},
have been experimentally realized in optical platforms~\cite{Zhaoscience,XuNatNano,FengNaturePhoton},
quantum walk systems~\cite{WangLaser},
cold atom systems~\cite{ZhouPRA}, circuit QED systems~\cite{Starkov,Huang},
PT-symmetric acoustic systems~\cite{Zhu}, trapped ion systems~\cite{CaoPRL,Ding},
graphene metamaterials~\cite{LiuOE}, time-mode-locked lasers~\cite{LeefmansNP}, and others.

We discuss two different schemes to realize the gate $G$: one with a two-mode many-boson system and the other with postselection.
Our analysis reveals that, while both schemes can be implemented in principle, they require an
exponentially large amount of physical resources. This suggests that the extraordinary computational power
of non-Hermitian quantum computers (NQC) originates from the vast physical resources needed
to implement the non-unitary gate $G$.  Although we have only examined these two schemes,
we anticipate that other potential implementations of $G$ would similarly demand exponentially large amounts of physical resources.

\section{Theoretical Model of NQC}

In this section, we introduce a quantum computing model that removes the unitarity constraint on gates,
which we refer to as non-Hermitian quantum computing (NQC). We provide an algorithm for solving
the maximum independent sets (MIS) problem to illustrate its potential.
By employing an approach similar to that used for the Lorentz quantum computer~\cite{ZhangWu},
we prove that the class of problems efficiently solvable with NQC, denoted as BNQP
(bounded-error non-Hermitian quantum polynomial time), is equivalent to \(\text{P}^{\sharp\text{P}}\).

Non-Hermitian quantum mechanics can be  either  a fundamental but hypothetical theory~\cite{Bender,He,ZhangNJP}
or an effective theory for open quantum systems~\cite{Gao}.  For the former approach,
one must address the issue of action at a distance brought by non-unitary evolution~\cite{Susskind}.
In this work, we adopt the latter perspective, viewing the non-Hermitian system as an effective outcome resulting
from specific treatments of Hermitian systems.  In the subsequent section, we will delve
into the realization of NQC with real physical systems. We will highlight the practical challenges
involved in surpassing current quantum computing models and emphasize
the intricate relationship between computation and physics.

\subsection{Universal quantum gates without the unitarity constraint and BNQP}

We consider a computational model consisting of qubits and linear logic gates, where the unitarity constraint
on the gates is relaxed. All qubits remain within the Hilbert space; that is, for a qubit
state \(|\psi\rangle=c_0|0\rangle+c_1|1\rangle\), the length of the state \(|\psi\rangle\) is still defined
as \(d=\sqrt{\langle\psi|\psi\rangle}\). However, with the unitarity constraint removed, \(|\psi\rangle\) may not be
normalized. Consequently, the probabilities of measuring the states \(|0\rangle\) and \(|1\rangle\) are
\(\frac{|c_0|^2}{d^2}\) and \(\frac{|c_1|^2}{d^2}\), respectively.

In traditional quantum computers, gates must be unitary. It is well known that universal logic gates with
the unitarity constraint consist of both single-bit gates $\{H, T\}$ and a two-bit controlled-NOT (CNOT) gate~\cite{Nielson}.
The single-bit gates are the Hadamard gate $H$ and the $\pi/8$ gate $T$, given by
\be
	H = \frac{1}{\sqrt{2}}\left(\sigma_{x}+\sigma_{z}\right), \quad T = e^{-i\frac{\pi}{8}}
    \left(\begin{array}{cc}
	e^{i\pi/8} & 0 \\
	0 & e^{-i\pi/8}
	\end{array}\right) \,.
\ee
These gates act on individual qubits, and their combined use can approximate any single-qubit transformation
with arbitrary precision. The CNOT gate operates such that the target qubit $|\phi\rangle$ flips
only when the control qubit $|\psi\rangle$ is in the state $|1\rangle$. The combination of $H$, $T$,
and CNOT can be used to implement any unitary quantum operation on the entire system with arbitrary precision.

For the model of non-Hermitian quantum computer (NQC), we introduce a non-unitary linear single-qubit gate operation
\be  \label{Phi}
G=\left(\begin{array}{cc}g^{-1} & 0 \\ 0 & g   \end{array}\right),
\ee
where $g>0$ and $g\neq1$ is a real number.  Applying the gate $G$ $r$ times, we have
\be  \label{Phir}
G^r\equiv \underbrace{G\cdot G \cdots G}_r=\left(\begin{array}{cc}g^{-r} & 0 \\ 0 & g^{r}   \end{array}\right).
\ee
For simplicity, we will denote the continuous application of $G$ as $G^r$ in the following.

{\bfseries Theorem 1 (Pure State Controllability).} The gates $H$, $T$, CNOT, and $G$ allow for pure state controllability in NQC. That is, for any two single-qubit states $|\psi_i\rangle$ and $|\psi_f\rangle$, there exists a sequence of these gates that maps $|\psi_i\rangle$ to $|\psi_f\rangle$, up to a possible change in norm. This does not imply full operator controllability (the ability to approximate arbitrary linear operators), but is sufficient for state preparation in our algorithmic constructions.

{\it\bfseries  Proof}: We first review the proof that $H$, $T$, and CNOT form a set of universal gates for conventional
quantum computer~\cite{Elementary1995,Nielson}. The proof consists of three main steps.
(1) For single-qubit gates, the combination of the $H$ and $T$ gates can simulate any arbitrary unitary operation.
This means that for any given single-qubit state, we can use an appropriate combination of $H$ and $T$ to transform
it into any desired state, preserving the same magnitude. (2) The arbitrary  two-qubit controlled unitary gate, C-$\mathbb{U}$,
can be implemented through an appropriate combination of the $H$, $T$, and CNOT gates. $\mathbb{U}$ is an arbitrary
unitary operation on a single qubit. (3) An arbitrary unitary operation on the entire multi-qubit computer can
be realized by an appropriate combination of arbitrary single-qubit unitary operations and arbitrary controlled
unitary operations C-$\mathbb{U}$.

In our proof that $H$, $T$, CNOT, and $G$ (assume without loss of generality that $g>1$) allow for pure state controllability
in NQC, the proof that $H$, $T$, and CNOT form a set of unitary universal gates
reviewed briefly above will be directly used as it can be found in any textbook on quantum computation~\cite{Nielson}.
Our proof has three steps.

The first step is to prove that $H$, $T$, and $G$ together can map any initial single-qubit state $|\psi_i\rangle$ to any desired final single-qubit state $|\psi_f\rangle$, with the caveat that
$B=\sqrt{\langle\psi_f|\psi_f\rangle}$ may not  equal to $A=\sqrt{\langle\psi_i|\psi_i\rangle}$.

We define the function $f(c_1,c_2)=\sqrt{|g^{-1}c_1|^2+|gc_2|^2}$. It is clear that for any  $c_1$ and $c_2$ satisfying
the relation $|c_1|^2+|c_2|^2=A^2$, we have  $g^{-1}A\leq f(c_1,c_2)\leq gA$.  With some algebra, one can
show that, as $c_1,c_2$ vary continuously with the constraint $|c_1|^2+|c_2|^2=A^2$,
the value of $f(c_1,c_2)$ continuously varies within the interval $\left[g^{-1}A,gA\right]$.
The proof then proceeds by considering three cases: $g^{-1}\leq B/A\leq g$, $B/A>g$, and $B/A<g^{-1}$.

(i) In the case of $g^{-1}\leq B/A\leq g$, we first use an appropriate combination of $H$ and $T$ to rotate $|\psi_i\rangle$
to some state  $(c_1,c_2)^T$  such that it satisfies $f(c_1,c_2)=B$.  We then apply the single-qubit gate $G$
to $(c_1,c_2)^T$ to obtain the state $(g^{-1}c_1,gc_2)^T$. Finally, we use an appropriate combination of $H$ and $T$
to rotate $(g^{-1}c_1,gc_2)^T$ to $|\psi_f\rangle$.

(ii) For the case of  $B/A>g$, we set $r=\text{Floor}\left[\frac{\ln(B/A)}{\ln g}\right]$, where $\text{Floor}[x]$ denotes
the greatest integer less than or equal to $x$. We first use an appropriate combination of $H$ and $T$ to
rotate  $|\psi_i\rangle$ to $(0,A)^T$. Then, we apply gate $G$ to $(0,A)^T$ $r$ times,
yielding the state $|\psi_i'\rangle=(0,g^rA)^T$, with its modulus denoted as $A'$. We now have $g^{-1}\leq B/A'\leq g$.
Finally, repeating the  step for the first case, we can rotate $|\psi_i'\rangle$ to $|\psi_f\rangle$.

(iii) When $B/A<g^{-1}$, we set $r=\text{Floor}\left[-\frac{\ln(B/A)}{\ln g}\right]$. We first use an
appropriate combination of $H$ and $T$ to rotate $|\psi_i\rangle$ to $(A,0)^T$. Then, we apply
gate $G$ to $(A,0)^T$ $r$ times, yielding the state $|\psi_i'\rangle=(g^{-r}A,0)^T$,
with its modulus denoted as $A''$. We now have $g^{-1}\leq B/A''\leq g$. Finally, repeating the
step for the first case, we can rotate $|\psi_i'\rangle$ to $|\psi_f\rangle$.
\begin{figure}[h!]	
\vspace{0.5cm}
\begin{subfigure}[b]{0.40\columnwidth}	
  		~~~~~\Qcircuit @C=1.2em @R=1.5em {
				\lstick{|\psi\rangle}   &   \ctrl{1}     &   \qw      \\
				 \lstick{|\phi\rangle}   &   \gate{G}   &    \qw       \\			 	
			 }  	
	\caption{}
\end{subfigure}
%\hspace{0.5cm}
\begin{subfigure}[b]{0.45\columnwidth}	
  		~~~~~\Qcircuit @C=1.2em @R=1.5em {
				\lstick{|\psi\rangle} & \qw & \gate{\sigma_x} &  \ctrl{1}     &    \qw       &      \ctrl{1} &   \gate{\sigma_x} &\qw    \\
				 \lstick{|\phi\rangle} & \qw & \qw &   \gate{\sigma_x}&  \gate{G}   &   \gate{\sigma_x}       &  \gate{G}	&\qw		
}	
	\caption{}
\end{subfigure}
\caption{(a) Two-bit logical controlled-$G$ (C-$G$) gate;
(b) A  method to implement the C-$G$ gate using CNOT and $G$ gates.  In this scheme,
the parameter $g'$ for the C-$G$ gate in (a) is related to the parameter $g$ of the $G$ gate in (b) as $g'=g^2$. }
	\label{f3}
\end{figure}

The second step is to prove that an appropriate combination of the gates $H$, $T$, CNOT, and $G$ can realize
any two-qubit non-unitary controlled gate C-$\mathbb{NU}$, where $\mathbb{NU}$ represents a non-unitary
single-qubit operation applied to the target qubit. We proceed in two steps.
(i) We can use the CNOT and $G$ gates to generate the C-$G$ gate, as shown in Fig.~\ref{f3},
where $|\psi\rangle$ is the control qubit and $|\phi\rangle$ is the target qubit. The single-qubit NOT gate $\sigma_x$
on the control qubit can be generated using $T$ and $H$. For a C-$G$ gate, the $G$ operation on
the target qubit $|\phi\rangle$ is applied only when the control qubit $|\psi\rangle = |1\rangle$.
(ii) As we can  construct any unitary controlled gate C-$\mathbb{U}$ using $H$, $T$, and CNOT~\cite{Nielson},
we can then construct any C-$\mathbb{NU}$ by combining C-$\mathbb{U}$ gates and C-$G$ gates.

Based on the proof for universal unitary gates, any arbitrary unitary transformation of a multi-qubit system
can be decomposed into a series of unitary single-qubit operations and two-qubit unitary controlled operations.
Similarly, any non-unitary operation on the entire system can be decomposed into a series of  non-unitary
single-qubit and two-qubit operations. Since the proof  is nearly identical to the unitary case, the details are not provided here.
\hfill   $\square$

Similar to the complexity class BQP defined for conventional quantum computer,
we now present the definition of the complexity class BNQP for NQC.

\textbf{\bfseries Definition of BNQP}. For a language L within BNQP, there exists a uniform family of quantum circuits, denoted as $\{\mathbb{C}_n\}_{n\geq1}$, where each circuit is of polynomial size in $n$. These circuits utilize qubits whose states reside in Hilbert space and employ universal gates without the unitarity constraint, such as $H$, $T$, CNOT, and $G$. They also allow measurements, after which no additional quantum gates can be applied. Given an input of length $n$ and specific initial states for the work qubits, the circuit $\mathbb{C}_n$ operates for polynomial time in $n$ and then halts. For $\omega\in\text L$, the probability of obtaining an accepting state is greater than $2/3$. Conversely, for $\omega\notin\text L$, this probability is less than $1/3$.

The criterion for an accepting state can involve either all bits in $\mathbb{C}_n$ or a single qubit. For example, a specific qubit, referred to as the ``Y qubit", can serve this purpose. An accepting state is defined as:
\be
|\text{accept}\rangle=|\Psi_1\rangle\otimes|1_Y\rangle,
\ee
while a rejecting state is defined as:
\be
|\text{reject}\rangle=|\Psi_2\rangle\otimes|0_Y\rangle,
\ee
where $|\Psi_1\rangle$ and $|\Psi_2\rangle$ represent arbitrary states of all qubits except for the ``Y qubit". The subscripts $Y$ in $|1_Y\rangle$ and $|0_Y\rangle$ denote the states of the ``Y qubit".

For $\omega\in\text L$, the output of $\mathbb{C}_n$ will be of the form:
\be \label{ZZZZ}
|\psi\rangle=c_{\text{yes}}|\Psi_1\rangle\otimes|1_Y\rangle+c_{\text{no}}|\Psi_2\rangle\otimes|0_Y\rangle,
\ee
where $\frac{|c_{\text{yes}}|^2}{|c_{\text{yes}}|^2+|c_{\text{no}}|^2}>2/3$.

Similarly, for an input $\omega\notin\text L$ of length $n$, the output of $\mathbb{C}_n$ will also be in the form of the expression above, but with $\frac{|c_{\text{yes}}|^2}{|c_{\text{yes}}|^2+|c_{\text{no}}|^2}<1/3$. For convenience, this error probability is often expressed as an exponentially small quantity rather than using $1/3$.  When we restrict the universal logic gates to $H$, $T$, and CNOT, the definition reverts to that of BQP.    \hfill   $\square$

\subsection{NQC algorithm for the maximum independent sets}

To demonstrate the remarkable computational power of NQC, we present an algorithm capable of solving the maximum independent sets (MIS) problem in polynomial time. Since the problem of finding an MIS in a graph composed of vertices and edges is NP-hard, the Cook-Levin theorem implies that NQC can, in principle, solve all NP problems in polynomial time; that is, NP $\subseteq$ BNQP.

For a graph $G(n,m)$ with $n$ vertices and $m$ edges, an independent set (IS) is a subset of the vertices that are not directly connected by edges. The MIS(s) are those with the largest number of vertices among all ISs. Finding the MIS(s) of a given graph is a challenging task on classical computers, as it is an NP-hard problem~\cite{Xiao}. A recently proposed quantum algorithm shows promising signs of exponential speedup~\cite{Yu2021,WYW}; however, there is no rigorous proof or very convincing numerical evidence. Here we present an NQC algorithm that can solve MIS problems in polynomial time.

To design the algorithm for a given graph $G(n,m)$, we assign a Boolean variable to each vertex, $x_1,x_2,\cdots,x_n$.  As a result, a subset of the vertices is represented by an integer $x$ in its $n$-digit binary form: if its $i$th digit $x_i=1$ then the $i$th vertex is in the subset; $x_i=0$ then it is not. If $x$ is an IS, then its $x_i$ and $x_j$ cannot both be $1$ simultaneously if the two vertices $x_i$ and $x_j$ are connected by an edge.

For an NQC algorithm, we use $n$ work qubits to represent the $n$ vertices. Their $N=2^n$ possible states $|00...0\rangle$, $|00...1\rangle$, ..., $|11...1\rangle$ naturally represent all the subsets of vertices. That is, a basis vector $\ket{x}$ corresponds to the subset $x$ where the integer $x$ is in its binary form.
The goal is to find the target state(s) $|M\rangle$ that corresponds to MIS(s) out of the $N=2^n$ possible states.

In our algorithm for the MIS problem, we introduce an additional qubit, denoted $|\ldots_n\rangle$, where the subscript ``n" indicates that this qubit may undergo non-unitary transformations, in parallel with the $n$ work qubits in the computational circuit.

An MIS must first be an IS. To distinguish between ISs and non-ISs, we use the following oracle,
\begin{equation} \label{misOracle}
O_{\rm IS}=(I-P_{\rm IS})\otimes I_o+P_{\rm IS}\otimes (|0_n\rangle\langle1_n|+|1_n\rangle\langle 0_n|)\,,
\end{equation}
where $I_o$ is the identity matrix for the oracle qubit and $P_{\rm IS}$ is a projection onto the sub-Hilbert space spanned by all possible solutions $|x\rangle$ of IS,
\be
P_{\rm IS}=\sum_{x\in {\rm IS}}|x\rangle\langle x|\,.
\ee
The quantum oracle $O_{\rm IS}$ is similar to the one used in the Grover algorithm~\cite{Nielson}, and it evaluates whether a subset $x$ is an IS in polynomial time.

As $g>1$, the circuit of our algorithm is shown in Fig.~\ref{t4}. The initial state of the whole system, including the $n$ work qubits and the additional qubit that may undergo non-unitary transformations, is set to be $|00\ldots0\rangle\otimes|0_n\rangle$. The algorithm then proceeds as follows:

\begin{figure}[t]
\hspace*{0.5cm}
	\centerline{
  \Qcircuit @C=0.35em @R=1.5em {
	 \lstick{\ket{0}} & \gate{H} & \qw& \qw  &   \multigate{4}{{\rm Oracle}}  & \qw&\ctrl{4}&\qw&\qw&\qw&\qw&\qw&\qw&\qw&\qw&\qw\\
	 \lstick{\ket{0}}  &  \gate{H} & \qw & \qw &   \ghost{{\rm Oracle}}  & \qw &  \qw&\qw&\ctrl{3}&\qw&\qw&\qw&\qw&\qw&\qw&\qw\\
	 \lstick{}   &    &  &  \lstick{\vdots~~}   &    &  && & &&&&&&&&\lstick{\cdot\cdot\cdot\cdot\cdot\cdot~~~~~}&\\
	 \lstick{\ket{0}}  &   \gate{H}   & \qw &\qw & \ghost{{\rm Oracle}}  & \qw  &  \qw&\qw
	 &\qw&\qw  &\qw &\qw&\qw
	 &\ctrl{1}&\qw &\qw
	 \inputgroupv{2}{3}{4.5em}{1.1em}{{\rm work~qubits:} ~~~~~~~~~~~~~~~~~~~~~~~~~~~~~}\\
     \lstick{\rm n-H\ qubit : \ket{0_n}} & \qw & \qw& \qw  &   \ghost{{\rm Oracle}}  & \qw&\gate{G^{r}}&\qw&\gate{G^{r}}\qw&\qw&\qw&\qw&\qw&\gate{G^{r}}&\qw&\qw\\
	   }
  }
\caption{Circuit of an NQC algorithm for solving the MIS problem, which is NP-hard.
	The big box represents the oracle that implements the operator ($\ref{misOracle}$).
	The state $|0_n\rangle$ with subscript $n$ stands for the state of the qubit that may undergo non-unitary transformation.
	The gate $H$ represents the Hadamard gate. The gate $G$ and $G^r$ is for the gate shown in Eqs.~(\ref{Phi}) and (\ref{Phir}). As explained in the text, $r$ is proportional to $n$. The controlled gates are $C-G^{r}$ to correctly implement the amplitude scaling in Eq. (11).}
	\label{t4}
\end{figure}

(i) Apply Hadamard gates to all work qubits;

(ii) Apply the oracle $O_{\rm IS}$;

(iii) For each work qubit as the control qubit and the additional qubit (which may undergo non-unitary transformations) as the target qubit, perform $n$ instances of the C-$G^{r}$ gate operation. (Since we have established the universality of non-unitary gates, the C-$G^{r}$ gate can be implemented using universal non-unitary gates);

(iv) Measure the additional qubit that may undergo non-unitary transformations, denoted $|\ldots_n\rangle$.

After the step (i), the state becomes
\begin{align}\label{iniH}
|\Psi_1\rangle=\frac{1}{\sqrt{N}}\left(\sum_{x=0}^{2^n-1}|x\rangle\right)\otimes|0_n\rangle\,.
\end{align}
With the oracle operation in the step (ii), we have
\begin{align} \label{10}
|\Psi_2\rangle=&O_{\rm IS}|\Psi_1\rangle
=\sum_{x\notin\text{IS}}|x\rangle\otimes|0_n\rangle
+\sum_{x\in\text{IS}}|x\rangle\otimes|1_n\rangle \,.
\end{align}
After the step (iii), we obtain
\begin{align} \label{result}
|\Psi_3\rangle=\sum_{x\notin\text{IS}}g^{-m(x)r}\ket{x}\otimes\ket{0_n}
+\sum_{x\in\text{IS}}g^{m(x)r}\ket{x}\otimes\ket{1_n},
\end{align}
where $m(x)$ is the number of ones in the binary form of $x$, or equivalently, the number of vertices in the IS $x$.
According to Eq.~(\ref{result}), the probability $P$ of obtaining the MIS(s) after measuring $|\ldots_n\rangle$ is given by
\begin{equation} \label{misp}
P=\frac{g^{2Mr}N_{\text{MIS}}}{\sum_{x\notin\text{IS}}g^{-2m(x)r}+\sum_{x\in\text{IS}}g^{2m(x)r}},
\end{equation}
where $M$ is the number of vertices in an MIS and $N_{\text{MIS}}$ is the number of  MIS(s). It is obvious that we have
\begin{equation} \label{13}
P>\frac{g^{2Mr}N_{\text{MIS}}}{g^{2(M-1)r}(N-N_{\text{MIS}})+g^{2Mr}N_{\text{MIS}}}.
\end{equation}
When $g>1$, it is clear that $P\approx1$ when $r\approx\frac{1}{\ln g}\ln{N}\propto n$. Since there are  $n$ C-$G^{r}$ gates, the time complexity of our algorithm is $O(n\ln{N})\sim O(n^2)$. If $g<1$, we only need to make a slight modification to the circuit to obtain the same result.

The definition of BNQP is intrinsically tied to decision problems. By trivially extending the circuit, we can transform the problem into a decision problem. The input for the decision algorithm is of the form ``$G(n,m)$ (a graph with $n$ vertices and $m$ edges) + $S$ (a subset of vertices in $G$)", and the output is of the form shown in Eq.~(\ref{ZZZZ}), where $\frac{|c_{\text{yes}}|^2}{|c_{\text{yes}}|^2+|c_{\text{no}}|^2}$ will be nearly $1$ if $S$ forms an MIS in $G$, and nearly $0$ if $S$ does not.

This extension to the decision circuit can be easily achieved by adding an additional oracle to the original circuit. The portion of input $G(n,m)$ is used for the initial circuit in Fig.~2, while the portion of $S$ serves as the input for the additional oracle. By checking whether the output of the circuit in Fig.~2 matches $S$ by the additional oracle, we can solve this decision problem. We provide a step-by-step explanation: (1) Run the circuit in Fig.~2 to obtain a candidate MIS. (2) Use an additional oracle to compare this candidate with $S$. (3) Output ``yes" if they match, ``no" otherwise.

In fact, the MIS problem is also in the ${\text P}^{\text{NP}}$ complexity class belonging to PH (polynomial hierarchy). The $k$-IS, which involves finding an independent set of $k$ vertices, falls within NP. We can submit $n$ non-adaptive queries to the NP-oracle (or SAT-oracle since the SAT problem is NP-complete) for $0$-IS, $1$-IS, all the way up to $n$-IS solutions. By determining the maximum value of $k_{\text{MAX}}$ that yields a positive result from the oracle, we derive the solution for MIS. Thus, the MIS problem is in the class ${\text P}^{\parallel\text{NP}}$​​~\cite{Buss,Hemachandra}.

Furthermore, based on Ref.~\cite{ZhangWu}, we can state the following:

{\bfseries Theorem 2}. The class of languages decidable in polynomial time by NQC is equivalent to the complexity class \pnp, i.e., $\text{BNQP}={\text P}^{\sharp \text{P}}$.

{\it\bfseries  Proof}: The equivalence $\text{BNQP} = {\text P}^{\sharp\text{P}}$ follows essentially the same reasoning as the proof for the Lorentz quantum computer (LQC) given in Ref.~\cite{ZhangWu}. However, since the gate sets in the two models are different, we clarify the mapping between them and explain why the same complexity result holds.

The LQC uses hyperbolic bits (hybits) and Lorentz transformations, with universal gates including the Hadamard gate $H$, the $\pi/8$ gate $T$, the $\tau$ gate for hybits, and controlled-$\sigma_z$ gates between qubits and hybits. In contrast, our NQC uses ordinary qubits throughout, but allows a non-unitary single-qubit gate $G$ Eq.~(\ref{Phi}) alongside the standard unitary gates $H$, $T$, and CNOT.

Despite these differences, both models possess the key capability of ``exponential amplitude amplification" in a controllable, state-dependent manner. In LQC, this amplification is achieved through the hyperbolic rotation $V$ (Eq.~(8) of Ref.~\cite{ZhangWu}) acting on hybits, enabled by the indefinite inner product and the unobservability of the $|1\rangle$ hybit state. In NQC, the same amplification effect is produced by the non-unitary gate $G^{r}$ (Eq.~(\ref{Phir})), which scales the amplitude of $|0\rangle$ by $g^{-r}$ and that of $|1\rangle$ by $g^{r}$.

The mapping between the two models in the context of the complexity proof is as follows:
\begin{itemize}
    \item The role of the hybit in LQC is played in NQC by an ordinary qubit that is subjected to the $G$ gate. The exponential amplification of the wavefunction of the hybit in the observable state $|0\rangle$ in LQC corresponds to the fact that in NQC, after applying $G^r$, the amplitude of $|1\rangle$ becomes exponentially amplified relative to that of $|0\rangle$.
    \item The controlled-hyperbolic gate CV in LQC (Fig.~(4) of Ref.~\cite{ZhangWu}) is replaced in NQC by the C-$G$ gate (Fig.~\ref{f3}), which can be constructed from CNOT and $G$ gates.
    \item The multi-controlled amplification used in LQC algorithms (e.g., the CCV gate) can be similarly realized in NQC using multiple C-$G$ gates, since we have shown that controlled non-unitary operations can be built from the universal set $\{H, T, \text{CNOT}, G\}$.
\end{itemize}

With this mapping, the algorithmic constructions in Ref.~\cite{ZhangWu}—such as those for solving MAJSAT and for demonstrating $\text{P}^{\sharp\text{P}} \subseteq \text{BLQP}$—can be directly translated into NQC circuits by replacing hybit operations with sequences of $G$ gates and their controlled versions. The state amplification and selection mechanisms that underpin the complexity proofs are preserved.

Therefore, although the physical implementations and gate sets differ, the computational capabilities of NQC and LQC are equivalent in terms of polynomial-time solvability. The proof that $\text{BLQP} = {\text P}^{\sharp\text{P}}$ in Ref.~\cite{ZhangWu} relies solely on the ability to perform exponential amplification and to implement controlled amplifications, both of which are present in NQC. Hence, we conclude that $\text{BNQP} = {\text P}^{\sharp\text{P}}$.    \hfill   $\square$

%%%%%%%%%%%%%%%%%%%%%%%%%
\section{Physical scheme to implement gate $G$}
%%%%%%%%%%%%%%%%%%%%%%%%%

Previous approaches to enhancing quantum computing capabilities, such as nonlinear schemes~\cite{Abrams}, utilizing closed timelike curves~\cite{Bacon}, exploiting the power of postselection~\cite{aaronson,knill}, and proposing the Lorentz quantum computer~\cite{He,ZhangWu}, all involve granting quantum computers hypothetical abilities. In contrast, our non-unitary logic gate scheme can be designed and analyzed from a physical perspective. Below, we present a proposal for a double-well cold atom system, where the particle number is not fixed, to introduce the non-Hermitian mechanism. Although we can always attempt to propose physical schemes for implementing non-Hermitian quantum computation, the difficulty is insurmountable. This provides us with an initial insight into the relationship between physics and computation: it is unrealistic to attempt to surpass quantum computation in an era where no new physics has been discovered.

\subsection{Implementing the single-qubit gate $G$ by altering the particle number}

Quantum computation utilizes qubits, which are two-level systems; therefore, we consider a two-level system in this context. To introduce a non-unitary gate, we examine a multi-particle two-level system. Specifically, we consider a single-component multi-particle bosonic cold atom double-well system (e.g., Rb atoms). When the particle number $N$ is sufficiently large, this system can be described by the Gross-Pitaevskii equation:
\be
\text{i}\hbar\frac{\partial}{\partial t} \left(\begin{array}{c}a\\b \end{array}\right)=H_{\text{GP}}\left(\begin{array}{c}a\\b \end{array}\right),
\ee
with the non-linear Gross-Pitaevskii Hamiltonian operator given by
\begin{equation} \label{gene-H0}
H_{\text{GP}}=\left(\begin{array}{cc}\epsilon_1+c|a|^2 &x-\text{i}y  \\x+\text{i}y &  \epsilon_2+c|b|^2 \end{array}\right).
\end{equation}
Here, the parameters $x$ and $y$ represent the tunneling strength between the two wells. As the barrier between the cold atoms in the wells increases, the value of $\sqrt{x^2+y^2}$ decreases. In the double-well potential formed by the optical field, the ratio of  $x$ and $y$ can be adjusted by changing the phase of the light. The parameter $\epsilon_1-\epsilon_2$ represents the energy bias between the two wells, and $c$ characterizes the interaction between the atoms. In the Gross-Pitaevskii equation, $|a|^2$ and $|b|^2$ represent the particle numbers in the left and right wells, respectively. Thus, the classical wavefunction $(a,b)^T$ corresponds to the Fock state $||a|^2,|b|^2\rangle$, where there are $|a|^2$ particles in the left well and $|b|^2$ particles in the right one.

Now, since the square of the modulus of a state $|\psi\rangle$ corresponds to the particle number associated with the state, we can define the length of the wave function as $d^2=\langle\psi|\psi\rangle$, ensuring that the wave function resides within a Hilbert space. It is important to note that this differs from the initial conception of quantum mechanics potentially involving intrinsic non-Hermitian elements. When non-Hermitian quantum mechanics was first proposed, the concept of an indefinite inner product was introduced~\cite{Dirac,Pauli}, where the length of a state $|\psi\rangle$ is defined as $d^2=\langle\psi|\eta|\psi\rangle$, with $\eta$ being a Hermitian matrix. Clearly, for a general $\eta$, $|\psi\rangle$ does not lie within a Hilbert space.

When the particle number is large but conserved, a normalized wavefunction $|a'|^2+|b'|^2=1$ is typically introduced, with $a'=a/\sqrt{N},b'=b/\sqrt{N}$. However, since we will introduce a gain and loss mechanism for the particles, we directly set $|a|^2+|b|^2=N$ instead, which allows us to reflect the increase and decrease in particle numbers.

To implement the non-unitary gate $G$, no inter-atomic interactions are needed, but a particle gain and loss mechanism must be introduced. For the first requirement，by tuning the magnetic field via Feshbach resonance~\cite{Chin}, we set the scattering length between the bosons to zero. For the second requirement, we guide ions (Rb$^+$) into the left well using an electromagnetic field, and then, through Coulomb interactions or electric field control, the ions can be captured and neutralized by the neutral cold atoms, forming new cold atoms. By carefully adjusting the injection rate, we ensure an exponential growth relationship, provided that transitions between the two wells are absent. In practice, if we use cold atoms in high Rydberg states for charge transfer, ionize them, and guide them into the left well via an electric field, the particle injection process can be precisely controlled by tuning the Rydberg state lifetime and ionization rate. In the other well (the right well), by applying controlled optical losses, we excite the particles in the right well to high-lying excited states and use another laser to induce their escape. By selecting the appropriate laser frequency, we make the particles in the right well resonate with the optical field, thus enhancing the loss rate. Alternatively, the right well can be coupled to an external environment, causing its evolution under the Lindblad form of the master equation to control the loss, ensuring an exponential decay.

Through the particle gain and loss mechanism, the effective Hamiltonian can be written as:
​\begin{equation} \label{gene-H}
H_{\text{GP}}=\left(\begin{array}{cc}\epsilon_1+\text{i}s &x-\text{i}y  \\x+\text{i}y &  \epsilon_2-\text{i}s \end{array}\right),
\end{equation}
where the parameter $s$ represents the strength of the gain and loss. This Hamiltonian has been recently studied to explore the topology of energy bands and the properties of edge states~\cite{T9,Fu,chen1,chen2}. For the implementation of a non-unitary gate $G$, it is sufficient to set the energy bias $\epsilon_1-\epsilon_2=0$, and we will only discuss this case moving forward. In fact, the Hamiltonian (\ref{gene-H}) with non-Hermitian term $\text{i}s\sigma_z$ is valid in the sense of single particle measurement, which will be discussed in the following.

The non-Hermitian Hamiltonian (\ref{gene-H}) has biorthogonal eigenstates $|\psi_0\rangle$, $|\psi_1\rangle$ and $|\phi_0\rangle$, $|\phi_1\rangle$, with $\langle\phi_i|\psi_j\rangle=\delta_{ij}$, which satisfy the following equations:
\ba \nonumber
&&H_{\text{GP}}|\psi_0\rangle=E_0|\psi_0\rangle,\quad H_{\text{GP}}|\psi_1\rangle=E_1|\psi_1\rangle, \\\nonumber
&&\langle\phi_0|H_{\text{GP}}=\langle\phi_0|E_0, \quad \langle\phi_1|H_{\text{GP}}=\langle\phi_1|E_1.
\ea
It is important to note that only the right states $|\psi_0\rangle$ and $|\psi_1\rangle$ have physical meaning, with  $\langle\psi_{0(1)}|\psi_{0(1)}\rangle$ representing the occupation number of state $0(1)$, while the left states $|\phi_0\rangle$ and $|\phi_1\rangle$ serve only an auxiliary role. This differs from the non-Hermitian mechanics where an indefinite inner product is introduced~\cite{Dirac,Pauli}. For non-Hermitian systems with an indefinite inner product, the left and right states are related by a Hermitian matrix $\eta$: $|\psi_{0(1)}\rangle=\eta|\phi_{0(1)}\rangle$, and the length of a state is defined as  $\langle\psi_{0(1)}|\eta|\psi_{0(1)}\rangle=\langle\phi_{0(1)}|\psi_{0(1)}\rangle$.

We define the states $|0\rangle\equiv|\psi_0\rangle$ and $|1\rangle\equiv|\psi_1\rangle$ for the discussion of quantum computation. When the parameters $x$ and $y$ are scanned slowly for a fixed $s$, we can realize the adiabatic evolution of the states  $|0\rangle$ and $|1\rangle$ along with the Berry phase. Through a dynamical analysis, we obtain a purely imaginary geometric phase. When the parameters $x$ and $y$ are controlled to undergo a loop evolution, we can realize the gate $G$. By performing $r$ loops, we can realize the gate $G^r$, as shown in Eqs.~(\ref{Phi}) and (\ref{Phir}). The explicit expressions for the left and right states, as well as the detailed analysis of the Berry phase, are provided in the Appendix A.

The concept of classical approximation, specifically the Gross-Pitaevskii equation, requires some clarification. When the effective Planck constant (which is $\hbar/N$ for a system of $N$ particles) tends to zero, the system’s dynamics will reduce to classical mechanics. However, at this point, the quantum mechanical principle of superposition is not violated, meaning that entanglement in the quantum system can still persist. The true mechanism behind the disappearance of superposition and entanglement is decoherence, which arises from the interaction between the system and its external environment. Therefore, even for a system with a large particle number $N$, in principle, the system can still be entangled with other quantum systems. This is exactly what we need for quantum computing.

When $N\rightarrow\infty$ and the normalized wavefunction $|a'|^2+|b'|^2=1$ (with $a'=a/\sqrt{N}$ and $b'=b/\sqrt{N}$) is used, any small change in $|a'|^2$ or $|b'|^2$ corresponds to a much larger change in the particle number. Therefore, for systems with a large particle number and the normalized condition, the Gross-Pitaevskii equation remains valid. However, when using $(a,b)^T$, under $N\rightarrow\infty$, the dynamics described by the Gross-Pitaevskii equation are only valid when the changes in $|a|^2$ and $|b|^2$ are sufficiently large (i.e., far greater than $1$ or unchanged). In cases where the changes in $|a|^2$ and $|b|^2$ are small, the Gross-Pitaevskii equation fails to provide an accurate description.

\subsection{The single-particle measurement in a two-mode many-boson system}

In fact, the non-Hermitian Hamiltonian is only valid in the sense of single-particle measurement; we here discuss it. To this end, we consider an open quantum system where each boson can be in one of two modes, $\ket{0}$ and $\ket{1}$. To realize the functionality of a qubit, we focus on the following state:
\be \label{DD0}
|\Psi\rangle=\frac{c_0}{\sqrt{N!}}(\hat{b}_0^\dag)^N|\text{Vac}\rangle+ \frac{c_1}{\sqrt{M!}}
(\hat{b}_1^\dag)^M|\text{Vac}\rangle.
\ee
where $\hat{b}_0^\dag|\text{Vac}\rangle=\ket{0}$ and $\hat{b}_1^\dag|\text{Vac}\rangle=\ket{1}$,
 $|c_0|^2+|c_1|^2=1$, $N$ and $M$ are two positive integers. This state can function as a qubit if we perform single-particle measurements. The probability of finding a boson in state $\ket{0}$ ($\ket{1}$) is proportional to $N|c_0|^2$ ($M|c_1|^2$). This can be seen from the one-particle reduced density matrix:
 \be
 \label{rho}
\hat{\rho}=\sum_{i,j\in\{0,1\}}|i\rangle\langle\Psi|\hat{b}_j^\dag\hat{b}_i|\Psi\rangle\langle j|
=\left(\begin{array}{cc}N|c_0|^2&0\\0&M|c_1|^2\end{array}\right)\,.
\ee
This type of single-particle measurement is conceptually similar to ARPES, where the emitted single electrons reveal the states of electrons in the material. It is clear that this many-boson system can function as a qubit.

In this physical implementation of the computing model, we have two types of qubits: conventional qubits, which are operated on by unitary gates, and the qubit realized by the aforementioned many-boson system, which is operated on by the non-unitary gate $G$. The gain and loss required for implementing $G$ are achieved by increasing the number of bosons in state $\ket{0}$ and decreasing the number of bosons in state $\ket{1}$, respectively.
Formally, we should have
\be \label{DD1}
G|\Psi\rangle=c_0 \frac{1}{\sqrt{(gN)!}}(\hat{b}_0^\dag)^{gN}|\text{Vac}\rangle+c_1 \frac{1}{\sqrt{(M/g)!}}
(\hat{b}_1^\dag)^{M/g}|\text{Vac}\rangle\,.
\ee
For simplicity, we assume that $g=2$ and $N=2^n,M=2^m$.

During or at the end of the computation, the whole system may get entangled via gate operations and becomes
\be \label{jieshi3}
\ket{\Phi}=c_0|x\rangle_q\otimes \frac{1}{\sqrt{N!}}(\hat{b}_0^\dag)^N|\text{Vac}\rangle
+c_1|y\rangle_q\otimes \frac{1}{\sqrt{M!}}(\hat{b}_1^\dag)^M|\text{Vac}\rangle\,,
\ee
where $|x\rangle_q$ and $|y\rangle_q$ ($\langle x|y\rangle=0$) are two distinct states of all the
qubits other than the qubit realized with the many-boson system.
Although this state is different from $\ket{\Psi}$, it yields the same single boson density
matrix $\rho$ in Eq.~(\ref{rho}). It means that after making a single-particle measurement,
if $N|c_0|^2\gg M|c_1|^2$, one would end with state $|x\rangle_q$ with great certainty.
If $|x\rangle_q$ is the solution we need to obtain, we have a successful algorithm.

Now, we examine the entangled state of a multi-particle system and a qubit:
\be \label{exa1}
|\varphi\rangle=c_0\left(\begin{array}{c}1\\0\end{array}\right)_q\otimes \left(\begin{array}{c}a\\0\end{array}\right)+c_1\left(\begin{array}{c}0\\1\end{array}\right)_q\otimes \left(\begin{array}{c}0\\b\end{array}\right),
\ee
where the subscript ``q" represents the qubit state, and $(a,0)^T$ and $(0,b)^T$ denote the Gross-Pitaevskii wavefunction of the multi-particle system. In this state, if we measure the qubit to determine which mode it is in, the qubit will collapse to  $|0\rangle\equiv(1,0)_q^T$ with probability $|c_0|^2$, and to $|1\rangle\equiv(0,1)_q^T$ with probability $|c_1|^2$.

Now, if we apply a $G$-gate as shown in Eq.~(\ref{Phi}) to the multi-particle system in the initial state $|\varphi\rangle$, the final state becomes:
\be \label{jieshi2}
|\varphi\rangle=c_0\left(\begin{array}{c}1\\0\end{array}\right)_q\otimes \left(\begin{array}{c}g a\\0\end{array}\right)+c_1\left(\begin{array}{c}0\\1\end{array}\right)_q\otimes \left(\begin{array}{c}0\\g^{-1}b\end{array}\right).
\ee
For the state in Eq.~(\ref{jieshi2}), if we measure the qubit or measure the multi-particle state in the particle number representation, the probability of the qubit collapsing to $|0\rangle$ with probability $|c_0|^2$ does not change. However, if we design an experimental setup to measure which well a particle in the multi-particle system is located in, e.g., by manipulating the double-well optical field to generate relative velocity between the two wells, allowing a particle to escape from the system and interact with the apparatus to detect its momentum, thus determining the original well, we will measure the probability of a single particle being in the left well to be proportional to $|gc_0|^2$, and the probability of it being in the right well to be proportional to $|c_1/g|^2$.

At this point, the change in the magnitude of the wavefunction can be treated directly as a c-number, i.e.,
\be \label{jieshi3b}
c_0\left(\begin{array}{c}1\\0\end{array}\right)\otimes \left(\begin{array}{c}g a\\0\end{array}\right)+c_1\left(\begin{array}{c}0\\1\end{array}\right)\otimes \left(\begin{array}{c}0\\g^{-1}b\end{array}\right)\sim gc_0\left(\begin{array}{c}1\\0\end{array}\right)\otimes \left(\begin{array}{c} a\\0\end{array}\right)+g^{-1}c_1\left(\begin{array}{c}0\\1\end{array}\right)\otimes \left(\begin{array}{c}0\\b\end{array}\right).
\ee
This provides a physical interpretation of using a wavefunction to describe the qubit and using $(a,b)^T$ to describe the multi-particle system. If we upgrade the entangled state with a single qubit to an entangled state with multiple qubits, the situation remains completely consistent. For a more detailed discussion on single-particle measurements in a multi-particle system, refer to Appendix B.

However, measuring the single-particle state in a multi-particle system is extremely difficult to implement in practice, which highlights a fundamental challenge in improving the performance of quantum computers from a physical perspective. In particular, in low-energy physical processes occurring in systems like cold atoms or condensed matter, realizing particle-number-uncertain states like in Eq.~(\ref{DD0}) is itself extremely challenging.

The above scheme, while physically motivated, presents significant technological challenges, with some steps requiring an exponentially large amount of physical resources.
\begin{itemize}
\item When $N=M$, the state in Eq.~(\ref{DD0}) is the generalized GHZ state. It is well known that
it is very difficult to prepare such a state in experiment. Although, in theory, a GHZ state with thousands
of atoms can be prepared using entanglement amplification~\cite{Zhaonpj}, in practical superconducting
systems, a GHZ state with $N=27$ has been prepared~\cite{Mooney}. A similar state with $N\neq M$ is more difficult to prepare.
\item To implement $G$ as in Eq.~(\ref{DD1}), it is not only
difficult but also requires exponentially large number of bosons.
As indicated in Eq.~(\ref{DD1}), after the operation of $G$,
the number of bosons pumped into state $\ket{0}$ is proportional to $N$ while
the number of bosons removed from state $\ket{1}$ is proportional to $M$.
This means that, in such an implementation,
 the gate $G$ depends on the state it operates on and this is in stark contrast with
 other conventional quantum gates whose operations are independent of the states that they act on.
 This makes the physical implementation of $G$ very difficult. Moreover,
 because of the gate $G$, the number of bosons can increase or decrease exponentially, and it requires that we have exponentially large number of bosons available.
 \item  It is clearly a challenge to realize the entangled state  $\ket{\Phi}$
 in Eq.~(\ref{jieshi3}). That it is difficult to prepare
 the state $\ket{\Psi}$ is because it is easy for $\ket{\Psi}$ to get entangled
 with the environment and experience decoherence.  However, to entangle with other qubits and achieve the state $\ket{\Phi}$, delicate manipulation of the systems is required.
\end{itemize}
The above analysis shows that it is a daunting challenge to implement
a qubit operated on by the non-unitary gate $G$ with a many-boson system.
It requires not only sophisticated techniques but also exponentially large number of bosons.

\subsection{The number of particles required for NQC: exponential}

It is necessary to briefly mention: how many particles are required for the single-particle measurement algorithm? The most efficient approach seems to be utilizing $g\gg1$, which makes $b/g\rightarrow0$ in Eq.~(\ref{jieshi2}). In this case, as long as $a\neq0$ and $c_0\neq0$, the measured state of the particle in a single-particle measurement will always be in the $0$ mode (left well), regardless of how small $|c_0|^2\neq0$ is. However, this choice would render the algorithm ineffective because, as discussed in the context of single-particle measurements, there is a significant probability that no particle will be detected.

For example, in the state shown in Eq.~(\ref{DD0}), where $M\neq0$ and $N\neq0$, based on our discussion, the probability of measuring the particle in the $|0\rangle$ state is proportional to $N|c_0|^2$, the probability of measuring the particle in the $|1\rangle$ state is proportional to $M|c_1|^2$, and the probability of detecting no particle is proportional to $(N-M)|c_1|^2$. In this case, when $|c_0|^2\neq0$, as long as $N\gg M$, the combined probability of measuring the single particle in the $|0\rangle$ state and detecting no particle approaches $\frac{N|c_0|^2+(N-M)|c_1|^2}{N|c_0|^2+(N-M)|c_1|^2+M|c_1|^2}\approx1$. When $|c_0|^2=0$, both the probabilities of measuring the single particle in the $|0\rangle$ state and detecting no particle are zero, and we can only measure the particle in the $|1\rangle$ state. That is, whether $|c_0|^2$ is zero can be completely distinguished through single-particle measurement. When $M=0$, corresponding to the vacuum state in the second term of Eqs.~(\ref{DD0}), (\ref{exa1}), and (\ref{jieshi2}), when $|c_0|^2\neq0$ and $N\gg M$, the combined probability of measuring the single particle in the $|0\rangle$ state and detecting no particle still approaches $1$. Moreover, when $|c_0|^2$ is very small, the probability of detecting no particle approaches $1$, and when $|c_0|^2=0$, no particle will be detected. Therefore, whether $|c_0|^2=0$ cannot be distinguished.

To ensure the algorithm is effective, it is necessary to have at least one particle in the state described by the second term of Eq.~(\ref{jieshi2}). Additionally, for the first term, the particle number must be significantly larger than $2^n$, where $n$ represents the length of the input to the problem. This means that the particle number must be at least exponential in terms of the input length. For example, in the MIS algorithm discussed in Sec. II, $n$ corresponds to the number of vertices, which is typically equivalent to the length of the input. According to Eq.~(\ref{misp}), the number of particles required should be much larger than $2^n$. This introduces another challenge: decoherence, which we will address later.

\subsection{Scheme to implement two-qubit control gate C-$G$}

In a double-well system, the most easily implementable two-qubit controlled gate is the CZ gate. Once the CZ gate is realized, the CNOT gate can also be implemented, and consequently, the controlled gate C-$G$ can be achieved. The definition of the CZ gate is as follows: when the control qubit is $|\psi\rangle=|1\rangle$, the operator $\sigma_z$ acts on the target qubit $|\phi\rangle$, while when the control qubit is $|0\rangle$, nothing happens. In the specific algorithm, we only need to perform non-unitary operations on certain qubits, such as the MIS algorithm discussed in Sec.~II. Similar to the Lorentz quantum computer~\cite{ZhangWu}, we can design analogous algorithms to solve the \pnp problem, where only one qubit needs to undergo a non-unitary operation. Thus, in this case, we only consider the scenario where the control qubit is carried by a single-particle double-well system and the target qubit is carried by a multi-particle double-well system capable of performing non-unitary operations. The case where both qubits are single-particle systems is simpler to implement. As for the case where the control qubit is a multi-particle system, this is not needed in practical algorithms.

\begin{figure}[t]
\begin{center}
\vspace*{-0.1cm}
\par
\resizebox *{10cm}{5cm}{\includegraphics*{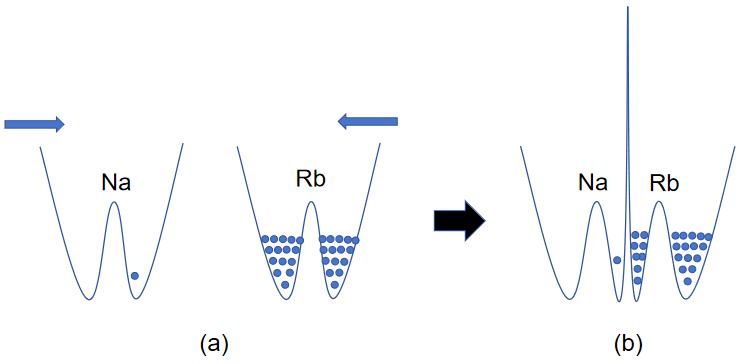}}
\end{center}
\vspace*{0cm}
\caption{(a) Traditional qubits are realized using a single-particle double-well system, while qubits capable of non-unitary operations are realized using a multi-particle double-well system. Using Feshbach resonance, we achieve zero inter-particle interactions in the multi-particle system, while particles in the multi-particle system interact with those in the single-particle system. (b) The CZ gate is implemented by bringing one well of the single-particle system close to one well of the multi-particle system. The potential barrier between the two double-well systems is sufficiently high, such that particles in each double-well system cannot tunnel into the other system. This CZ gate, as shown in the Appendix A, can be directly interpreted as a CNOT gate in the computational basis.}
\label{CZ}
\end{figure}

When performing the CZ gate operation, as shown in Fig.~\ref{CZ}, we can bring two double-well systems close enough so that a particle in one well of a double-well interacts with a particle in another well of a different double-well. The barrier between the two wells is sufficiently high, ensuring that no particles can tunnel between the two double-wells.

For example, in a multi-particle double-well system, we use Rb atoms, while for the single-particle case, we use Na atoms. By utilizing Feshbach resonance, we can adjust the external field to make the interaction between Rb atoms vanish, while interactions between Na and Rb atoms still exist.

Assume that the energy biases of the two double-well systems are both zero. The Hamiltonian of the system is:
\be
H=(x_1-\text{i}y_1)\hat{c}_1^\dag \hat{c}_2+(x_1+\text{i}y_1)\hat{c}_1\hat{c}_2^\dag+(x_2-\text{i}y_2)\hat{a}^\dag \hat{b}+(x_2+\text{i}y_2)\hat{a}\hat{b}^\dag+V\hat{c}_2^\dag\hat{c}_2\hat{a}^\dag\hat{a},
\ee
Here, $(x_1,y_1)$ ($(x_2,y_2)$) are the tunneling amplitudes for the Na (Rb) double wells, $V$ represents the interaction strength between each Rb atom in the left well and the Na atom in the right well, and $(\hat{c}_1,\hat{c}_2)$ ($(\hat{a},\hat{b})$) are the operators for Na (Rb) atoms in the (left well, right well) configuration. When the number of Rb atoms is sufficiently large, we approximate $a$ and $b$ as c-numbers to introduce the Gross-Pitaevskii equation. In this case, the effective Hamiltonian for the Gross-Pitaevskii wavefunction $(a,b)^T$ related to the Rb double-well system can be written as:
​\begin{equation} \label{gene-H2}
H_{\text{GP}}=x_2\sigma_x+y_2\sigma_y+\text{diag}(V\hat{c}_2^\dag\hat{c}_2,0)=\left(\begin{array}{cc}V\hat{c}_2^\dag\hat{c}_2&x_2-\text{i}y_2\\
x_2+\text{i}y_2& 0 \end{array}\right)    \,.
\end{equation}

It is evident that when all other effects are turned off, including the particle's gain and loss, as well as the tunneling of the particle in the double-well, the effective Gross-Pitaevskii Hamiltonian operator becomes,
​\begin{equation} \label{gene-H3}
H_{\text{GP}}=\left(\begin{array}{cc}V\hat{c}_2^\dag\hat{c}_2&0\\
0&0 \end{array}\right)    \,.
\end{equation}
In this case, no interaction occurs when the Na atom is in a well far from the Rb system (the left well, $|0\rangle$ state for the Na double-well, where the eigenvalue of $\hat{c}_2^\dag\hat{c}_2$ is $0$). However, when the Na atom is in a well near the Rb system (the right well, $|1\rangle$ state for the Na double-well, the eigenvalue of $\hat{c}_2^\dag\hat{c}_2$ is $1$), the Gross-Pitaevskii wavefunction in one of the Rb wells (left well) acquire a phase $-Vt/\hbar$ due to the interaction, where $t$ is time. By choosing the time $t=\hbar\pi/V$, we can implement the CZ gate (the overall phase $e^{\text{i}\pi}=-1$ is always irrelevant to the universal gate, so we typically ignore it when discussing universal gates). Afterward, the two double-well systems are separated. The CZ gate here is defined with respect to the double-well representation of the multi-particle system, where $(a,0)^T$ ($(0,b)^T$) represents the particle being in the left (right) well. However, in the representation used for implementing the $G$ gate, the CZ gate here can be directly considered as the CNOT gate, as detailed in the Appendix A.

It is important to note that we cannot treat the single-atom values $\hat{c}_1$ and $\hat{c}_2$ as c-numbers. Although any component of the wavefunction describing a single atom can be treated as a c-number, this is fundamentally different in physical meaning from the classical approximation when the particle number is sufficiently large. In the single-particle double-well system, after measurement, the particle can only collapse into either the left or right well with a probability determined by the wavefunction.

Once the CZ gate is available, the CNOT gate can be implemented using the Hadamard gate $H$, as shown in Fig.~\ref{CX}. The Hadamard gate can be realized by adjusting the height of the barrier between the double-wells and the energy bias between the two wells. From the discussion of universal gates earlier, we can then implement the C-$G$ gate.

\begin{figure}[h!]	
\vspace{0.5cm}
\begin{subfigure}[b]{0.45\columnwidth}	
  		~~~~~\Qcircuit @C=1.2em @R=1.5em {
				\lstick{|\psi\rangle} &     \qw       &      \ctrl{1} &  \qw &\qw    \\
				 \lstick{|\phi\rangle} &   \gate{H}   &   \gate{\sigma_z}       &  \gate{H}	&\qw		
}	
	
\end{subfigure}
\caption{Using the CZ and Hadamard gates $H$ to realize the CNOT gate.}
	\label{CX}
\end{figure}

\subsection{Gain and loss realized with postselection}

Beside using the change of particle numbers, one can
also use postselection to  implement gain and loss in quantum systems.
As we will see below, one also faces a barrier of exponential scale.

In this scheme, all the qubits are the same. To implement $G$ on any one
of the qubits, one introduces an auxiliary qubit. The target qubit is initially in
an arbitrary state of $\alpha\ket{0_n}+\beta\ket{1_n}$, and the
auxiliary qubit is in the state of $\ket{0}$, i.e., the state for these two qubits
is $\ket{\psi_0}=(\alpha\ket{0_n}+\beta\ket{1_n})\otimes \ket{0}$.
We apply the following unitary operation on them~\cite{Chen} (with $0<\eta<1$ a real constant),
\be \label{Ugate}
U=\left(
\begin{array}{cccc}
1 & 0 &0 & 0\\
0& \eta & \sqrt{1-\eta^2} & 0\\
0& -\sqrt{1-\eta^2} & \eta & 0\\
0 & 0 &0 & 1
\end{array}
\right)
\ee
and obtain
\be
\ket{\psi_1}=\alpha\ket{0_n}\otimes|0\rangle+\beta\eta\ket{1_n}\otimes|0\rangle+\beta\sqrt{1-\eta^2}\ket{0_n}\otimes|1\rangle\,.
\ee
After a measurement on the auxiliary, if the result is $\ket{0}$, the target
qubit is in the following state
\be
\ket{\varphi_1}=\frac{1}{\sqrt{|\alpha|^2+\eta^2|\beta|^2}}(\alpha\ket{0_n}+\beta\eta\ket{1_n})\,.
\ee
This is equivalent to applying $G$ to $\ket{\varphi_0}$ with $g=\eta^{1/2}<1$, showing how gain and loss of the gate $G$ can be implemented on a qubit by postselecting $\ket{0}$ on the auxiliary qubit. Redefining the states fully realizes the case $g>1$.
This is similar to the dilation method used to observe parity-time symmetry breaking in a single nitrogen-vacancy center in diamond~\cite{nitrogen}. Such a postselection has also been discussed in the context of measurement with non-Hermitian systems~\cite{Yang}.

Consider the task of solving NP-complete problems (e.g., SAT) or NP-hard problems such as MIS. In such cases, we encounter states of the form given in Eq.~\eqref{10}. In the extreme case, where the amplitude associated with $|0_n\rangle$ is exponentially small in $n$, this state can be expressed as (here, for convenience, we swap the roles of $|0_n\rangle$ and $|1_n\rangle$ compared to Eq.~\eqref{10})
\be
c_1|\varphi_0\rangle\otimes|0_n\rangle+ c_2|\varphi_1\rangle\otimes|1_n\rangle,
\ee
where $|c_1|^2\sim d^{-n}$, $|c_2|^2\sim 1-d^{-n}$ with $d>1$, and $|\varphi_0\rangle$ and $|\varphi_1\rangle$ are normalized states of the work qubits satisfying $\langle\varphi_0|\varphi_1\rangle=0$.

After applying the gate $G$ $r$ times—in the MIS algorithm, we actually use $C$-$G^r$, which selectively operates on specific states within $|\varphi_0\rangle$ and $|\varphi_1\rangle$; however, the analysis using $G^r$ alone suffices to demonstrate the exponential decay of the postselection success probability, as the underlying mechanism remains identical—the state evolves to
\be
\frac{1}{\sqrt{|c_1|^2+|c_2|^2\eta^{2r}}}\left(c_1|\varphi_0\rangle\otimes|0_n\rangle+ \eta^r c_2|\varphi_1\rangle\otimes|1_n\rangle\right),
\ee
and according to Eq.~\eqref{Ugate}, the probability that all $r$ postselection steps succeed is given by
\be \label{POO}
P=\prod_{j=1}^r \frac{|c_1|^2+|c_2|^2\eta^{2j}}{|c_1|^2+|c_2|^2\eta^{2j}+|c_2|^2\eta^{2j-2}(1-\eta^2)}=|c_1|^2+|c_2|^2\eta^{2r}\sim d^{-n}+(1-d^{-n})\eta^{2r}.
\ee

As indicated by Eq.~\eqref{13}, the minimal requirement for the algorithm is $r\sim n$. Accordingly, the $n$-dependence of the success probability is given by:：
\be
P\sim d^{-n}+(1-d^{-n})\eta^{2n}.
\ee
Since $0<\eta<1$ and $d>1$, the success probability remains exponentially small in $n$. Thus, to guarantee the success of the algorithm, one must construct an exponentially large number of identical quantum computers running in parallel, so that at least one of them collapses the auxiliary qubit to $\ket{0}$ (i.e., succeeds). This clearly implies an exponential growth in physical resources with the input length $n$.

\subsection{Decoherence time analysis}

The primary challenge in realizing quantum computers lies in the fact that the decoherence time is shorter than the time required for the algorithm to run. The NQC suffers from an even shorter decoherence time than traditional quantum computers, making it more difficult to implement.

What we need to analyze is the decoherence induced by the interaction between the double-well system and the external environment. In the system we use, each neutral atom is uncharged, so the influence of external fields is negligible. The primary source of decoherence comes from impurities that collide with the cold atoms. In optical lattice cold atom experiments, there are always some sparse external particles diffusing through the system. Therefore, short-range scattering results in decoherence. Previous studies have shown that, in the position representation, the evolution of the single-particle density matrix over time, due to short-range scattering with environmental particles, is governed by~\cite{Tegmark2000}
\be
\rho(x,x',t)=\rho(x,x',0)f(x,x',t),
\ee
where
\be  \label{ff}
f(x,x',t)=\exp[-\Lambda t(1-e^{-|x'-x|^2/2\lambda^2})]\approx \left\{\begin{array}{lr}\exp[-|x'-x|^2\Lambda t/2\lambda^2]& \text{for} \, |x'-x|\ll\lambda \\ & \\e^{-\Lambda t} & \text{for} \, |x'-x|\gg\lambda   \end{array}     \right.
\ee
Here, $x$ and $x'$ represent spatial coordinates, $\lambda$ is the de Broglie wavelength of the atoms in the system, and $\Lambda=n\langle\sigma v\rangle$, where $n$ is the density of scatterers, $\sigma$ is the scattering cross-section, and $v$ is the relative velocity between the atoms and scatterers. The notation $\langle\ldots\rangle$ indicates the expectation value.

In the double-well system, we can approximate $|x-x'|$ as the distance between the two wells. In this case, the decay of $f(x,x',t)$ leads to the decay of the coherence between the two wells, causing the system to decohere. In cold atom systems, the temperature is typically extremely low, and the de Broglie wavelength $\lambda$ of the atoms is particularly long. Therefore, according to Eq.~(\ref{ff}), the decoherence time of the single-particle system is approximately:
\be
T_2=\frac{2\lambda^2}{|x-x'|^2\Lambda}=\frac{2\lambda^2}{|x-x'|^2n\langle\sigma v\rangle},
\ee
where $\lambda$ is large, so the decoherence time for the single-particle system can be quite long, which is an advantage of cold atom systems. Furthermore, the decoherence time depends on the concentration $n$ of impurities in the system that can scatter with the atoms, the scattering cross-section $\sigma$ reflecting the interaction strength, and the relative velocity between the atoms and impurities during the tunneling process. However, in any case, we can increase the decoherence time by reducing the concentration of impurities in the system through advanced techniques.

However, what we need now is a multi-particle system, and our algorithm requires the number of particles to increase exponentially with the problem input size (refer to Eq.~(\ref{misp}) in the MIS algorithm). For a system with $N$ particles, the decoherence time decreases compared to the single-particle system, as shown in~\cite{Tegmark2000,Schlosshauer}:
\be
T_2'=\frac{T_2}{N}.
\ee
The shorter decoherence time makes the implementation of non-unitary logic gate algorithms using multi-particle systems physically more challenging than in traditional quantum computers. This requires advanced technology to exponentially reduce impurity concentrations and the interaction between the system and its environment. Moreover, our system's gain and loss mechanism necessitates continuous interaction with the external environment. As such, in addition to requiring more advanced techniques for impurity reduction, we must ensure that interactions between the particles and the environment, beyond the gain and loss mechanism, do not introduce additional sources of decoherence. This highlights another fundamental challenge in enhancing the performance of quantum computers from a physical perspective.

\section{Summary}
In summary, from a practical perspective, rather than attributing hypothetical capabilities
to quantum computers or modifying quantum mechanics to enhance quantum computational power,
we focus on utilizing the gain and loss mechanisms of bosons to realize non-Hermitian quantum computation.
Theoretically, the corresponding complexity class, called bounded-error non-Hermitian quantum polynomial-time,
is equivalent to \pnp. However, from a practical standpoint, we face insurmountable challenges
to realize the gain and loss mechanism in experiment. For the two schemes that we have examined,
both require exponentially large amount of physical resources.
We expect that other schemes will be similarly high-demanding.
Such a burden is similar
in spirit to the exponential demand on classical memory when a classical computer is used to
simulate quantum many-body systems.

Our work unveils a fascinating and fundamental relationship between computational power and physics.
To go beyond quantum computing, it appears that we either need to endow quantum computers with
hypothetical capabilities or have to face insurmountable experimental barriers. The underlying mechanisms
behind this phenomenon will require further investigation in future research.

Finally, we make an observation. All the enhanced computing models~\cite{Deutsch,Bacon,Abrams,aaronson,knill,He,ZhangWu} and our current model can efficiently solve problems beyond the complexity class of NP. We can not help wonder whether it is possible to propose a  computing model by adding new physical principles so that the corresponding complexity class is equivalent to NP. As a result, the relation between these major complexity classes may be mapped into the relation between different sets of physical principles.

Note: After completing this work, we became aware of a related study by Barch and Lidar~\cite{Lidar} on non-Hermitian quantum computing. Both papers examine the computational power of non-Hermitian quantum dynamics and identify postselection as the essential mechanism underlying its apparent strength. The key difference lies in their focus and framing: Barch and Lidar prove a general complexity-theoretic equivalence—any non-unitary gate that is polynomially far from unitarity suffices to enable PostBQP, thereby establishing hardness under standard complexity conjectures. The present study, by contrast, constructs an explicit NQC model centered on a concrete gate $G$ and an algorithm for the maximum independent set problem, and then demonstrates through two physical implementation schemes that the model’s power collapses under exponential resource costs. This provides a direct, physically grounded trade-off between computational advantage and physical overhead. Despite subtle differences in theoretical approach, both analyses reach a similar conclusion: “there is no NH system which can be implemented efficiently as part of universal quantum computation”~\cite{Lidar}.

\section*{ACKNOWLEDGMENTS}
QZ is supported by the National Natural Science Foundation of China (Grant No. 12475018); 		
BW is supported by the National Natural Science Foundation of China (Grants No.  92365202,
No. 12475011, and No.  11921005),  the  National Key R\&D Program of China (2024YFA1409002),
and Shanghai Municipal Science and  Technology Major Project (Grant No.2019SHZDZX01).
QZ and BW are supported by Shanghai Municipal Science and Technology Project (Grant No.25LZ2601100).

\appendix

\section{Implementation of the $G^r$ gate by non-Hermitian geometric phase}

As $x^2+y^2>s^2$ and $\epsilon_1-\epsilon_2=0$, the biorthogonal eigenstates of the Hamiltonian (\ref{gene-H}) satisfying
$H_{\text{GP}}|\psi_0\rangle=E_0|\psi_0\rangle$, $H_{\text{GP}}|\psi_1\rangle=E_1|\psi_1\rangle$, $H_{\text{GP}}|\phi_0\rangle=E_0^*|\phi_0\rangle$, $H_{\text{GP}}|\phi_1\rangle=E_1^*|\phi_1\rangle$, and $\langle\phi_i|\psi_j\rangle=\delta_{ij}$ can be work out as
\ba  \nonumber \label{wf1}
|\psi_0\rangle=\left(\begin{array}{c} \sqrt{x^2+y^2-s^2}+is \\ \\ x+\text{i}y  \end{array}\right), \quad
|\phi_0\rangle=\left(\begin{array}{c} \frac{1}{2\sqrt{x^2+y^2-s^2}} \\ \\
\frac{\sqrt{x^2+y^2-s^2}+\text{i}s}{2(x-\text{i}y)\sqrt{x^2+y^2-s^2}} \end{array}\right), \\
|\psi_1\rangle=\left(\begin{array}{c} -x+\text{i}y \\ \\ \sqrt{x^2+y^2-s^2}+is \end{array}\right), \quad
|\phi_1\rangle=\left(\begin{array}{c}  -\frac{\sqrt{x^2+y^2-s^2}+\text{i}s}{2(x+\text{i}y)\sqrt{x^2+y^2-s^2}}\\ \\ \frac{1}{2\sqrt{x^2+y^2-s^2}}
 \end{array}\right),
\ea
with corresponding eigenenergies given by
\begin{equation} \label{energy}
E_{0(1)}=\pm\sqrt{x^2+y^2-s^2}\,.
\end{equation}
The eigenstates described in Eq.~(\ref{wf1}) are unique only up to a gauge freedom associated with $GL(1,\mathbb{C})$. It is important to emphasize again that the ket vectors $|\phi_0\rangle$ and $|\phi_1\rangle$ here do not have physical significance and are merely auxiliary tools. This is because, in the non-Hermitian system realized through the particle gain and loss mechanism, the square of the norm of the bra vectors $\langle\psi_{0(1)}|\psi_{0(1)}\rangle$ represents the particle number, meaning that the bra vectors still reside in the Hilbert space.

In the parameter space, the condition $x^2+y^2=s^2$ corresponds to the energy level degeneracy region. For non-Hermitian systems, the Hamiltonian cannot be diagonalized in this degenerate region, and thus the degenerate points are referred to as exceptional points. Unlike Hermitian operators, the right eigenvectors and left eigenvectors of a non-Hermitian Hamiltonian are generally different (though in special cases they may be the same).

When the parameters $\mathbf{R}=(x,y)$ of the Hamiltonian $H_{\text{GP}}$ vary with time, the eigenstates $|\psi_j(t)\rangle$ and eigenvalues $E_j(t)$ (for $j=0$ and $j=1$) become time-dependent. In this case, we can express the dynamical evolution as:
\begin{equation} \label{super}
|\psi(t)\rangle=\sum_j c_j(t) \exp\left[-\frac{\text{i}}{\hbar}\int_0^tE_j(t')dt'\right]|\psi_j(t)\rangle.
\end{equation}
Substituting this into the Schr\"odinger equation $\text{i}\hbar\partial|\psi\rangle/\partial t=H_{\text{GP}}|\psi\rangle$, and using the condition for the eigenstates, we obtain:
\be   \label{6}
\text{i}\hbar\sum_j\dot{c}_j(t)\exp[-\frac{\text{i}}{\hbar}\int_0^tE_n(t')dt']|\psi_j(t)\rangle
+\text{i}\hbar\sum_jc_j(t)\exp[-\frac{\text{i}}{\hbar}\int_0^tE_j(t')dt']|\dot{\psi}_j(t)\rangle=0.
\ee
Multiplying both sides of this equation by the left eigenstate $\langle \phi_m(t)|$, we get:
\be \label{7}
\dot{c}_m=-c_m\langle\phi_m|\dot{\psi}_m\rangle
-\sum_{j\neq m}c_j\langle\phi_m|\dot{\psi}_j\rangle \exp\left[-\frac{\text{i}}{\hbar}\int_0^t(E_j(t')-E_m(t'))dt'\right].
\ee
Assuming the system starts in the state $|\psi_m\rangle$, and invoking the adiabatic theorem, we expect that
$c_m\sim 1$ and $|c_j|\ll 1$ $(j\neq m)$ throughout the evolution. When all the eigenvalues $E_j$'s are real, the second term on the right-hand side can be neglected as,
\begin{equation} \label{condi}
\left| \frac{\hbar\langle\phi_m|\dot{\psi}_j\rangle}{E_m-E_j} \right|\ll 1, \text{for all} \; j\neq m.
\end{equation}
This condition can be derived by integrating Eq.~(\ref{7}). We then obtain the following simplified expression:
\be \label{adiasolution}
\dot{c}_m=-c_m\langle\phi_m|\dot{\psi}_m\rangle.
\ee
This is analogous to the situation in Hermitian systems. However, if the eigenvalues $E_j$ are complex, the second term in the equation can grow exponentially and cannot be neglected. In this case, the adiabatic theorem no longer holds.

Now, we are ready to derive the geometric phase. Suppose the system is initially in the state $\ket{\psi_j}$. If the adiabatic theorem holds, the system should evolve with time as:
\begin{equation} \label{e26}
|\psi(\mathbf{R})\rangle=|\psi_j(\mathbf{R})\rangle e^{-{\text i}\frac{\int E_j(\mathbf{R}) dt}{\hbar}}e^{{\text i}\beta_j}\,,
\end{equation}
where $\beta_j$ is the geometric phase. According to the result in Eq. (\ref{adiasolution}), the Berry connection can be expressed as~\cite{GarrisonPLA,GePRA,zhangPRA},
\begin{equation} \label{Berry-conn}
\mathbf{A}_j=\frac{\partial \beta_j}{\partial \mathbf{R}}=\text{i}\langle \phi_j(\mathbf{R})|\frac{\partial}{\partial \mathbf{R}}|\psi_j(\mathbf{R})\rangle.
\end{equation}
Since $|\psi_j\rangle$ is typically not equal to $|\phi_j\rangle$ for non-Hermitian Hamiltonians, the Berry phase is generally not real for non-Hermitian systems, even when all $E_j$'s values are real.

Using the Berry connection (\ref{Berry-conn}), the Berry curvature for the biorthogonal states, as shown in Eq.~(\ref{wf1}), can be directly obtained as,
\begin{eqnarray}  \nonumber \label{Berry-cur3}
 B_{0}&=& \frac{\partial A_{0y}}{\partial x}-\frac{\partial A_{0x}}{\partial y} =-\frac{\text{i}}{2}\frac{s}{(x^2+y^2-s^2)^\frac{3}{2}}, \\
 B_{1}&=& \frac{\partial A_{1y}}{\partial x}-\frac{\partial A_{1x}}{\partial y} =\frac{\text{i}}{2}\frac{s}{(x^2+y^2-s^2)^\frac{3}{2}}.
\end{eqnarray}
The purely imaginary Berry curvature is perpendicular to the $(x,y)$ plane, and diverges on the exceptional point
ring defined by $x^2+y^2=s^2$.

In the context of quantum computation, it is meaningful to utilize the eigenstates of the Hamiltonian
\be  \label{nqubit}
|0\rangle_n\equiv|\psi_0\rangle, \quad |1\rangle_n\equiv|\psi_1\rangle
\ee
as the two work states of a qubit on which non-unitary gates may act. As before, since the non-unitary gate acts on this qubit, we use the subscript $n$ to explicitly denote it for clarity.

In the $x-y$ parameter space, we can select a closed path outside the circle $x^2+y^2=s^2$ to control the parameters $x$ and $y$ adiabatically, as shown in Fig.~\ref{t10}, depending on the controllable range and convenience of the specific system. We can perform multiple loops along this path. When $s>0$, according to Eq.~(\ref{Berry-cur3}), the accumulated geometric phase from counterclockwise rotation along the selected path allows the norm of quantum amplitude of the state $|0\rangle_n$ to increase exponentially with the number of loops controlled along the path, while that of the state $|1\rangle_n$ decreases exponentially with the same number of loops. The situation for $s<0$ is exactly the opposite of that for $s>0$. The case of clockwise manipulation of parameters is also the reverse of the counterclockwise case. To avoid ambiguity, we define the gate $G$ based on the effect of one complete counterclockwise manipulation around a selected loop when $s>0$ (or equivalently, clockwise manipulation when $s<0$).
\begin{figure}[ht]
\setlength{\unitlength}{2mm}
	\begin{subfigure}[b]{0.45\columnwidth}
     \begin{picture}(25,25)
     \put(4,1){\vector(0,1){21}}
     \put(-7,10){\vector(1,0){38}}
     \put(4.5,21){$y$}
     \put(30,8.5){$x$}
     \put(4.3,8.5){$O$}
     \put(4,10){\vector(1,1){2.5}}
	\put(4,10){\circle{15}}
    \put(5,12){$s$}
    \qbezier(15,10)(15,15)(20,15)
    \qbezier(20,15)(24,15)(26,8)
    \qbezier(26,8)(26,5)(22,5)
    \qbezier(22,5)(15,5)(15,10)
    \put(15,17){a control loop}
      \end{picture}
	\end{subfigure}	
	\caption{The degenerate (exceptional points) circle $x^2+y^2=s^2$ for the Hamiltonian (\ref{gene-H}) and a suitable loop in the $x-y$ parameter space for adiabatic control to generate a purely imaginary geometric phase to realize the logic gate $G$ and $G^r$. By manipulating parameters $(x,y)$ along the loop for one full revolution, $G$ is implemented; for $r$ revolutions, $G^r$ is achieved.}
	\label{t10}
\end{figure}

Thus, the gate $G$ in the representation $(|0\rangle_n,|1\rangle_n)$ is precisely the expression shown in Eq.~(\ref{Phi}), given by
\be  \nonumber
G=\left(\begin{array}{cc}g&0\\0&g^{-1}   \end{array}\right),
\ee
where $g=e^{\text{i}\int_{\Gamma}B_{0}(x,y)dxdy}$. Here, $\ln g$ is a real quantity that represents the purely imaginary magnetic flux of the Berry curvature associated with the state $|0\rangle_n$, multiplied by $\text{i}$, as it traverses the control loop. The path $\Gamma$ represents this loop.

It should be noted in passing that, in the computational basis $(|0\rangle_n,|1\rangle_n)$, based on the eigenstate expression in Eq.~(\ref{wf1}), if we choose the $x$-coordinate of the initial point (which is also the final point) of the loop manipulation in Fig.~\ref{t10} to be zero, then according to Eq.~(\ref{wf1}), the CZ gate implemented in Eq.~(\ref{gene-H2}) is actually equivalent to the CNOT gate when $x^2+y^2>s^2$.

The effect of the gate $G^r$ can be expressed as:
\ba \nonumber|\psi(\text{f})\rangle_n&=&G^r|\psi(\text{i})\rangle_n=G^r(c_0|0\rangle_n+c_1|1\rangle_n)\\&=&g^rc_0|0\rangle_n+g^{-r}c_1|1\rangle_n. \ea
When the initial state $|\psi(\text{i})\rangle_n$ has non-zero amplitudes $c_1$ and $c_2$, the action of gate $G^r$ leads to a probability of measuring the state $|0\rangle_n$ in the final state $|\psi(\text{f})\rangle_n$ given by: $\frac{|g^rc_0|^2}{|g^rc_0|^2+|g^{-r}c_1|^2}\simeq 1$, while the probability of measuring the state $|1\rangle_n$ is:
$\frac{|g^{-r}c_1|^2}{|g^rc_0|^2+|g^{-r}c_1|^2}\simeq 0$. This holds for sufficiently large $r$ (and since the increase and decrease are exponential, $r$ typically does not need to be very large).

In non-Hermitian systems, one might consider designing Hamiltonians with complex eigenvalues to achieve exponential growth or decay of amplitude probabilities through time evolution. However, such exponential changes are extremely difficult to control in quantum circuits, as they occur at moments when an increase or decrease in amplitude is not needed. In contrast, choosing a purely imaginary adiabatic geometric phase to achieve exponential growth or decay of amplitude probabilities allows for more straightforward control of the quantum circuit, making the effective realization of NQC possible.

\section{Single-particle measurement in a multi-boson system}

\textbf{Case 1}. The situation with a fixed total particle number.

Let's consider a multi-particle state:
\be
\frac{1}{\sqrt{M!N!}}(b_0^\dag)^N(b_1^\dag)^M|\text{Vac}\rangle,
\ee
where $|\text{Vac}\rangle$ is the vacuum state, $b_0^\dag$ is the creation operator for the $|0\rangle$ state boson, and $b_1^\dag$ is the creation operator for the $|1\rangle$ state boson. In terms of the single-particle states $|0\rangle\equiv b_0^\dag|\text{Vac}\rangle$ and $|1\rangle\equiv b_1^\dag|\text{Vac}\rangle$, the above state can be written as:
\be
|\psi\rangle=\sqrt{\frac{M!N!}{(M+N)!}}\left[|0\rangle_1\otimes\ldots\otimes|0\rangle_N\otimes|1\rangle_{N+1}\otimes\ldots\otimes|1\rangle_{N+M}    +\text{other terms}\right],
\ee
where $|0\rangle_m$ denotes that the $m$th particle is in the $|0\rangle$ state, and $|1\rangle_m$ indicates it is in the $|1\rangle$ state. The term ``other terms" refers to the direct product state of $M+N$ particles, with $N$ particles in the $|0\rangle$ state and $M$ particles in the $|1\rangle$ state, excluding the given term. Suppose we measure in the particle number representation, i.e., the number of particles in the $|0\rangle$ state and the $|1\rangle$ state. In this case, the measurement will definitely yield $N$ particles in the $|0\rangle$ state and $M$ particle in the $|1\rangle$ state. However, if we perform a single-particle measurement, for example, by randomly ejecting one of the $M+N$ particles and keeping its state unchanged, then measuring its state, or using an experimental setup that can only measure one particle at a time, the probabilities are as follows:

If the particle indexed by ``$N+M$" is measured, the probability of finding it in the $|0\rangle$ state is $\frac{N}{M+N}$, and the probability of finding it in the $|1\rangle$ state is $\frac{M}{M+N}$. If the particle is measured to be in the $|0\rangle_{N+M}$ state, the wavefunction collapses to:
\be
|\psi\rangle=\sqrt{\frac{M!(N-1)!}{(M+N-1)!}}\left[|0\rangle_1\otimes\ldots\otimes|0\rangle_{N-1}\otimes|1\rangle_{N}\otimes\ldots\otimes|1\rangle_{N+M-1}    +\text{other terms}\right]\otimes|0\rangle_{N+M},
\ee
where the term ``other terms" refers to the direct product state of $M+N-1$ particles, with $N-1$ particles in the $|0\rangle$ state and $M$ particles in the $|1\rangle$ state, excluding the given term. Thus, the total symmetry of the $N+M$th particle is broken, while the total symmetry of the remaining $N+M-1$ particles is preserved.

If the particle is measured to be in the $|1\rangle_{N+M}$ state, the wavefunction collapses to:
\be
|\psi\rangle=\sqrt{\frac{(M-1)!N!}{(M+N-1)!}}\left[|0\rangle_1\otimes\ldots\otimes|0\rangle_{N}\otimes|1\rangle_{N+1}\otimes\ldots\otimes|1\rangle_{N+M-1}    +\text{other terms}\right]\otimes|1\rangle_{N+M},
\ee
where the term ``other terms" refers to the direct product state of $M+N-1$ particles, with $N$ particles in the $|0\rangle$ state and $M-1$ particles in the $|1\rangle$ state, excluding the given term.

We do not necessarily measure which of the $M+N$ particles it is, but the probability of measuring any particle in the $|0\rangle$ state is $\frac{N}{M+N}$, and the probability of measuring it in the $|1\rangle$ state is $\frac{M}{M+N}$. This shows that in a single-particle measurement, the probability of a particle being in a given state is proportional to the number of particles in that state.

For a superposition state with a fixed particle number, such as
\be \label{C1}
c_0\frac{1}{\sqrt{N!}}(b_0^\dag)^N|\text{Vac}\rangle+c_1\frac{1}{\sqrt{N!}}(b_1^\dag)^N|\text{Vac}\rangle,
\ee
its expression with the single-particle wavefunction is
\be
c_0|0\rangle_1\otimes|0\rangle_2\otimes\ldots\otimes|0\rangle_N+c_1|1\rangle_1\otimes|1\rangle_2\otimes\ldots\otimes|1\rangle_N.
\ee
The collapse of the single-particle measurement wavefunction and the resulting measurement can be obtained using a completely same method.

\textbf{Case 2}. The situation with an uncertain total particle number.

When the particle number is not fixed, for example, consider a multi-particle state with an uncertain number of particles (assuming  $N>M$):
\be \label{DD}
c_0 \frac{1}{\sqrt{N!}}(b_0^\dag)^N|\text{Vac}\rangle+c_1 \frac{1}{\sqrt{M!}}(b_1^\dag)^M|\text{Vac}\rangle.
\ee
Assuming $N>M$, then the state (\ref{DD}) can be rewritten in terms of single-particle states $|0\rangle\equiv b_0^\dag|\text{Vac}\rangle$ and $|1\rangle\equiv b_1^\dag|\text{Vac}\rangle$ as:
\be \label{DAN2}
|\psi\rangle=c_0 |0\rangle_1\otimes |0\rangle_2\otimes\ldots\otimes |0\rangle_N+c_1 |1\rangle_1\otimes |1\rangle_2\otimes\ldots\otimes |1\rangle_M,
\ee
where the subscripts $1,2,\ldots N$ represent different particles. The single-particle subscripts $\{1,2\ldots M\}$ in the second term are included within the $N$ particle subscripts $\{1,2,\ldots N\}$ of the first term because this represents an $N$-particle state. Due to the indistinguishability of the particles, it is entirely equivalent to replace the subscripts $\{1,2,\ldots M\}$ with any $M$ subscripts from $\{1,2,\ldots N\}$. Even when one or more of the subscripts $\{1,2,\ldots M\}$ are replaced by subscripts greater than $N$, there is no discernible difference between this and the state given in Eq.~(\ref{DAN2}), whether a single-particle or a Fock state measurement is performed.

In the case of a single-particle measurement on the state (\ref{DAN2}), if the particle indexed by $s$, with $M<s\leq N$, is measured, it will be found in the $|0\rangle$ state with probability $|c_0|^2$, and nothing will be detected with probability $|c_1|^2$. If the particle indexed by $s$, with $1\leq s\leq M$, is measured, it will be found in the $|0\rangle$ state with probability $|c_0|^2$, and in the $|1\rangle$ state with probability $|c_1|^2$. Since we do not know which particle is being measured, and due to the indistinguishability of particles, each particle contributes equally to the measurement.
Thus, after performing the single-particle measurement, the probability of detecting the particle in the $|0\rangle$ state is proportional to $N|c_0|^2$, while the probability of detecting it in the $|1\rangle$ state is proportional to $M|c_1|^2$. The probability of detecting nothing is proportional to $(N-M)|c_0|^2$.

By combining Case 1 and Case 2, we can derive the results of single-particle state measurements for other complex multi-boson states, where both the particle number and the quantum amplitude jointly determine the probability of the measurement outcome.

\end{CJK*}
\end{document}